\documentstyle[12pt]{article}
\catcode`\@=11 
\def\lsim{\mathrel{\mathpalette\@versim<}}
\def\gsim{\mathrel{\mathpalette\@versim>}}
\def\@versim#1#2{\vcenter{\offinterlineskip
    \ialign{$\m@th#1\hfil##\hfil$\crcr#2\crcr\sim\crcr } }}
\begin{document}
\begin{flushright}
CERN-TH.6999/93\\
CTP-TAMU-48/93\\
ACT-18/93
\end{flushright}
\vglue 0.5cm
\begin{center}
{\Large\bf Scrutinizing Supergravity Models through Neutrino Telescopes\\}
\vglue 1cm
RAJ GANDHI$^{(a),(b)}$, JORGE L. LOPEZ$^{(a),(b)}$, D.V.
NANOPOULOS$^{(a),(b),(c)}$,\\
KAJIA YUAN$^{(a),(b)}$, and A. ZICHICHI$^{(d)}$\\
\vglue 0.4cm
{\em $^{(a)}$Center for Theoretical Physics, Department of Physics,
Texas A\&M University\\}
{\em College Station, TX 77843-4242, USA\\}
{\em $^{(b)}$Astroparticle Physics Group, Houston Advanced Research Center
(HARC)\\}
{\em The Woodlands, TX 77381, USA\\}
{\em $^{(c)}$ CERN Theory Division, 1211 Geneva 23, Switzerland\\}
{\em $^{(d)}$ CERN, Geneva, Switzerland\\}
\vglue 1cm
{\tenrm ABSTRACT}
\end{center}

\noindent Galactic halo neutralinos ($\chi$) captured by the Sun or Earth
produce high-energy neutrinos as end-products of various annihilation modes.
These neutrinos can travel from the Sun or Earth cores to the neighborhood
of underground detectors (``neutrino telescopes") where they can interact
and produce upwardly-moving muons. We compute these muon fluxes in the
context of the minimal $SU(5)$ supergravity model, and the no-scale and
dilaton $SU(5)\times U(1)$ supergravity models. At present, with the
Kamiokande 90\% C.L. upper limits on the flux, only a small fraction of
the parameter space of the $SU(5)\times U(1)$ models is accessible for
$m_\chi\sim m_{\rm Fe}$, which in turn implies constraints for the
lightest chargino mass around 100 GeV for a range of $\tan\beta$ values.
We also delineate the regions of parameter space that would be accessible
with the improvements of experimental sensitivity expected in the near future
at Gran Sasso, Super-Kamiokande, and other facilities such as DUMAND and
AMANDA, currently under construction. We conclude that if neutralinos are
present in the halo, then this technique can be used to eventually explore
more than half of the allowed parameter space of these specific models, and
more generally of a large class of supergravity models, in many ways
surpassing the reach of traditional collider experiments.
\bigskip
\begin{flushleft}
\baselineskip=12pt
{CERN-TH.6999/93}\\
{CTP-TAMU-48/93}\\
{ACT-18/93}\\
September 1993
\end{flushleft}
\vfill\eject
\setcounter{page}{1}
\pagestyle{plain}

\baselineskip=14pt

\section{Introduction}
\label{sec:intro}
Even though there appear to be several indications that low-energy
supersymmetry is indeed a symmetry of Nature \cite{Haber}, no more converts
should be expected until an actual sparticle is observed experimentally or
unequivocal indirect experimental evidence becomes compelling. For either
discovery mode to be efficient, one must provide as accurate experimental
predictions as possible. In supersymmetric models where the number of
parameters is large, as in the minimal supersymmetric standard model (MSSM),
easily falsifiable experimental predictions are hard to obtain since the
various parameters usually allow a myriad of possibilities. One basic step
forward is to study specific supersymmetric models wherein the assumptions made
are well motivated by physics at very high energies, such as grand unification,
supergravity, and superstrings. In this way one can hope to test general
theoretical frameworks experimentally, as opposed to just ruling out small
regions of a many-dimensional parameter space. This line of thought has been in
existence for almost as long as supersymmetric phenomenology has. However, it
has only been since the firm establishment of the standard model at LEP that
real candidates for the theory beyond the standard model have emerged from a
morass of pre-LEP challengers.

The aim of this paper is to study a very promising
indirect experimental procedure to discover supersymmetry, in the context
of a class of $SU(5)$ and $SU(5)\times U(1)$ supergravity models, via
interactions of upwardly-moving muons in underground detectors.
As we will show, this method of experimental exploration is quite competitive
with other indirect probes which have been recently investigated in the
context of this same class of models, namely supersymmetric contributions to
the one-loop electroweak LEP observables \cite{ewcorr}, to the rare radiative
$b\to s\gamma$ decay \cite{bsgamma}, and to the anomalous magnetic moment
of the muon \cite{g-2}.

The basic idea is that the neutralinos $\chi$ (lightest linear combination of
the superpartners of the photon, $Z$-boson, and neutral Higgs bosons), which
are weakly interacting massive particles (WIMPs), are assumed to make up the
dark matter in the galactic halo, and can be gravitationally captured by
the Sun or Earth \cite{PS85,Gould}, after losing a substantial amount of
energy through elastic collisions with nuclei. The neutralinos captured in
the Sun or Earth cores annihilate into all possible ordinary particles,
and the cascade decays of these particles as well as their interactions with
the solar or terrestrial media produce high-energy neutrinos as
one of several end-products. These neutrinos can then travel from the
Sun or Earth cores to the vicinity of underground detectors, and interact
with the rock underneath producing detectable upwardly-moving muons
\footnote{In this paper, we do not consider the less promising ``contained
events'' in which neutrino interactions take place within the detector,
since the event rate for such events is proportional to $E_\nu$,
as opposed to $E^2_\nu$ for the upwardly-moving muon events.}.
Such detectors are rightfully called ``neutrino telescopes", and the
possibility of indirectly detecting various WIMP candidates has been
considered in the past by several authors \cite{others}.
More recent analyses can be found
in Refs.~\cite{RS,GR89,GGR91,KAM,KEK,Bott,FH}.

Making use of the techniques in the literature \cite{RS,GGR91,KAM},
our current work focuses on the predictions
for the upwardly-moving muon event rates (fluxes)
in two distinct supersymmetric models, namely,
the minimal $SU(5)$ supergravity model
and the string-inspired $SU(5)\times U(1)$ supergravity
model. In these models, because of the underlying structure of
supergravity, the low-energy couplings and masses are completely
determined via renormalization group equations (RGEs)
in terms of only five parameters: the three universal
soft-supersymmetry-breaking parameters ($m_{1/2},m_0,A$),
the top-quark mass ($m_t$), and the ratio of Higgs vacuum expectation
values ($\tan\beta$). The restriction to just five parameters is possible
because of the radiative electroweak symmetry breaking mechanism, which allows
to determine some further parameters, like the magnitude of the
Higgs mixing term $\mu$ (but not its sign). This mechanism imposes
further constraints on the model parameters ({\em e.g.},
it requires $\tan\beta>1$) and involves the value of
$m_t$ in a fundamental way \cite{aspects}.
These features make it possible to incorporate all currently available
experimental constraints, both direct and indirect, as well as
some theoretical consistency conditions into a complete determination
of the allowed parameter space of these models. We calculate the
upwardly-moving muon fluxes induced by the neutrinos from the Sun
and Earth in the still-allowed parameter space of these models,
and compare them with the currently most stringent 90\% C.L.
experimental upper bounds, obtained at Kamiokande,
for neutrinos from the Sun \cite{KEKII} and Earth \cite{KEK} respectively,
{\it i.e.}
\begin{eqnarray}
\Gamma_{\rm Sun}&<&6.6\times10^{-14} {\rm cm}^{-2} {\rm s}^{-1}
=2.08\times10^{-2}{\rm m}^{-2}{\rm yr}^{-1},\label{eq:Sup}\\
\Gamma_{\rm Earth}&<&4.0\times10^{-14} {\rm cm}^{-2} {\rm s}^{-1}
=1.26\times10^{-2}{\rm m}^{-2}{\rm yr}^{-1}.\label{eq:Eup}
\end{eqnarray}
Aiming at the next generation of underground experimental facilities,
such as MACRO and other detectors at the Gran Sasso Laboratory \cite{mac},
Super-Kamiokande \cite{sk}, DUMAND, and AMANDA \cite{amd}, where
improvements in sensitivity by a factor of 2--100 are expected,
we also delineate the region of the parameter space of these models
that would become accessible with an improvement of experimental
sensitivity by modest factors of two and twelve.

This paper is organized as follows. In Sec.~\ref{sec:model} we
describe the minimal $SU(5)$ and $SU(5)\times U(1)$ supergravity models.
In Sec.~\ref{sec:cap} we outline the calculation of the capture rates
at the Sun and Earth, while in Sec.~\ref{sec:det} we outline the
calculation of the corresponding detection rates.
In Sec.~\ref{sec:result} we present the results of our computations for the
two supergravity models described in Sec.~\ref{sec:model}.
Finally, we conclude in Sec.~\ref{sec:conclusion} with some comments.

\section{The Supergravity Models}
\label{sec:model}
We work in the context of  two supergravity models
based on the gauge groups $SU(5)$ \cite{Dickreview} and $SU(5)\times U(1)$
(``flipped $SU(5)$") \cite{EriceDec92}. In these two models, spontaneous
supersymmetry breaking takes place in the ``hidden'' sector, which manifests
itself in the ``observable'' sector Lagrangian as a set of
soft-supersymmetry-breaking terms that are universal at the respective
unification scales of the models. The low-energy particle content of these two
models is the same as that of the MSSM. However, mainly because of the
different gauge group structure, these two models have rather different
low-energy phenomenologies, which we now briefly describe in turn.

In the minimal $SU(5)$ supergravity model, the unification of the
standard model gauge couplings takes place at a scale
$M_U\sim10^{16}\,{\rm GeV}$ \cite{EKN}, as a combined result
of the $SU(5)$ gauge group and the minimal matter content.
This property of the model has been shown to hold even after the
most general low-energy and high-energy threshold corrections have
been incorporated, as well as the two-loop expressions for the RGEs
\cite{EriceDec92}. The unified symmetry implies the existence of
additional heavy particles with masses of order $M_U$. In particular, the Higgs
sector includes at least the {\bf24} representation to break $SU(5)$ down to
$SU(3)\times SU(2)\times U(1)$,
as well as the {\bf5}, ${\bar{\bf5}}$ superfields, whose scalar
doublet components ($H_2$) are responsible for the electroweak
gauge symmetry breaking and should be kept light, while
the scalar triplet components ($H_3$) must be heavy to avoid fast
proton decay through dimension-six operators.
It also follows that the Yukawa couplings of
the bottom-quark and tau-lepton must be unified at $M_U$,
which when evolved to low energies entails a constraint on the
relation between the ratio of Higgs vacuum expectation
values ($\tan\beta$) and the top-quark mass ($m_t$).
Perhaps the most distinguishing feature in the minimal $SU(5)$ model
has to do with the genuinely supersymmetric proton decay via
dimension-five operators, mediated by the exchange of superpartners
of the heavy Higgs triplets (${\tilde H}_3$).
Even for sufficiently large values of $M_{{\tilde H}_3}$,
it is necessary to tune the sparticle spectrum so that this type
of proton decay remains at an acceptable level \cite{ANpd}.
This usually requires light charginos and neutralinos, while squarks and
sleptons should be heavy. We study a generic supersymmetry breaking
scenario characterized by the universal parameters ($m_{1/2},m_0,A$).
It turns out that, because of the
correlation between the relic density of the lightest neutralino
and the soft supersymmetry breaking patterns \cite{susydm},
the cosmological constraint together with the proton decay
constraint dramatically reduce the parameter space of the minimal
$SU(5)$ model, and what remains are essentially points corresponding
to the resonances in the neutralino pair annihilation cross section
where $m_\chi\sim {1\over 2}M_Z,{1\over 2}m_h$ \cite{troubles}.
In addition, the ratio $\xi_0=m_0/m_{1/2}$ must be significantly
larger than unity, and $\tan\beta$ must not be too large
($\tan\beta\lsim3-5$) \cite{ANpd,LNPZ}. The ensuing simple relations
among the different sparticle masses in this model can be
summarized as:
\begin{equation}
{\rm Minimal}\ SU(5):\left\{
\begin{array}{ll}
{\rm Higgs:}&60\,{\rm GeV}<m_h<125\,{\rm GeV},\\
\chi^0_{1,2},\chi^\pm_1:&2m_{\chi^0_1}\sim m_{\chi^0_2}\sim m_{\chi^\pm_1}
\sim0.3 m_{\tilde g},\\
{\rm Squarks:}&m_{\tilde q}\approx\frac{1}{3}m_{\tilde g}\sqrt{6+\xi^2_0},\\
{\rm Sleptons:}&
m_{\tilde e_R}\approx\frac{1}{3}m_{\tilde g}\sqrt{0.15+\xi^2_0},
\quad m_{\tilde e_L}\approx\frac{1}{3}m_{\tilde g}\sqrt{0.5+\xi^2_0},
\end{array}
\right.
\label{eq:minmasses}
\end{equation}
where $\xi_0\gsim3$--$6$ depending on the value of the triplet higgsino
mass used ($M_{{\tilde H}_3}<(3$--$10)M_U$).
The actual five-dimensional parameter
space to be explored in this paper for the minimal $SU(5)$ model has
been obtained in Ref.~\cite{LNPZ}.

In the string-inspired $SU(5)\times U(1)$
supergravity model, the preferred unification scale is
$M_U\sim10^{18}\,{\rm GeV}$, as expected in the context
of string theory \cite{Lacaze}, which has been
realized by the effects of extra vector-like matter fields
at some intermediate scales \cite{Search}.
Contrary to the minimal $SU(5)$ model, the Higgs bosons needed to break
the $SU(5)\times U(1)$ symmetry belong to
the {\bf10} and $\overline{\bf10}$ representations, which can be
easily accommodated by the simplest string models (with Kac-Moody
level $k=1$), while the adjoint representation ({\bf24})
may only appear in more complicated constructions.
Because of the $SU(5)\times U(1)$ gauge group structure,
the doublet-triplet splitting of the pentaplet Higgs superfields
occurs in a very simple way, and the dimension-five proton decay
operators are automatically suppressed. In the case of the
$SU(5)\times U(1)$ supergravity model, we study two
string-inspired supersymmetry breaking scenarios:
(i) the no-scale model \cite{LNZI}, where $m_0=A=0$ \cite{LN}, and (ii) the
dilaton model \cite{LNZII},
where $m_0=\frac{1}{\sqrt{3}}m_{1/2},A=-m_{1/2}$ \cite{dilaton}.
Therefore, the $SU(5)\times U(1)$ model (either scenario) is quite
predictive since it depends on only three parameters:
$m_t,\tan\beta,m_{1/2}$. Also, in these two scenarios the
cosmological constraint is automatically satisfied \cite{susydm}.
The mass relations in this model are:
\begin{equation}
SU(5)\times U(1):\left\{
\begin{array}{ll}
{\rm Higgs:}&60\,{\rm GeV}<m_h<125\,{\rm GeV},\\
\chi^0_{1,2},\chi^\pm_1:&2m_{\chi^0_1}\approx m_{\chi^0_2}
\approx m_{\chi^\pm_1}\approx0.28 m_{\tilde g},\\
{\rm Squarks:}&m_{\tilde q}\approx m_{\tilde g},\\
{\rm Sleptons:}&m_{\tilde e_R}\approx 0.18(0.33)m_{\tilde g},
\quad m_{\tilde e_L}\approx 0.30(0.41)m_{\tilde g},
\end{array}
\right.
\label{eq:flpmasses}
\end{equation}
for the no-scale (dilaton) case. Note that the squark and slepton mass
relations in Eq.~(\ref{eq:minmasses}) do not reduce to those in
Eq.~(\ref{eq:flpmasses}), for values of $\xi_0=0$ (no-scale)
and $\xi_0=1/\sqrt{3}$ (dilaton), because of slight changes in the running of
the scalar masses down to low energies from the different starting
value of $M_U$. The allowed three-dimensional parameter space of this
model has been determined in Refs.~\cite{LNZI,LNZII}
for the no-scale and dilaton cases respectively.

We now comment on some other features about these two models
which are particularly relevant to our current work. First, for
$m_{\tilde g}\lsim 1$ TeV the lightest neutralino can not be a pure higgsino
\cite{ADD1}. In fact, we have found that for almost all the points in the
allowed parameter spaces of both models, the
lightest neutralino is a ``mixed'' state with substantial gaugino
and higgsino components. In the MSSM, such ``mixed'' neutralinos
normally cannot even account for the halo dark matter \cite{ADD2},
because the efficient annihilation via Higgs boson exchange
renders the relic density rather small ($\Omega_\chi h^2_0 < 0.05$).
This situation is improved in the two supergravity models,
as a result of the one-loop radiative corrections to the masses
of Higgs bosons which we have included in our analysis. Since
the one-loop corrected Higgs masses are normally larger than
the tree-level values, the overall effect is an enhancement of
the relic density for ``mixed'' neutralinos. Therefore,
the lightest neutralinos in these two models, although mostly
``mixed'' states, are still good candidates for the major component of
the galactic halo. Such neutralinos are mainly captured by the Sun and
Earth through their coherent (spin-independent) scattering off nuclei
due to the exchange of Higgs bosons. Furthermore, in these two models
the neutralinos typically have masses in the $20$--$150$ GeV range,
so their capture by the Earth is expected to be enhanced when the
neutralino mass closely matches that of the abundant elements in
the Earth's core (Fe) and mantle (Si and Mg), while the same effect
is irrelevant for the capture by the Sun \cite{Gould}.
We will discuss the implications of these features
in Sec.~\ref{sec:result}.

\section{The Capture Rate}
\label{sec:cap}

In order to calculate the expected rate of neutrino production due to
neutralino annihilation, it is necessary to first evaluate the rates at
which the neutralinos are captured in the Sun and Earth.
Following the early work of Press and Spergel \cite{PS85},
the capture of WIMPs by a massive body was studied extensively
by Gould \cite{Gould}. In this paper, we make use of Gould's formula,
and follow a procedure similar to that of Refs.~\cite{GGR91,KAM}
in calculating the capture rate.

{}From Eq.~(A10) of the second paper in Ref.~\cite{Gould}, the
capture rate of a neutralino of mass $m_\chi$ by the Sun or Earth
can be written as
\begin{equation}
C=\left({2\over 3\pi}\right)^{1\over 2}M_B
{\rho_\chi{\bar v}_\chi\over m_\chi}
\sum_i{{f_i\over m_i}\sigma_i X_i},\label{eq:nc}
\end{equation}
where $M_B$ is the mass of the Sun or Earth,
$\rho_\chi$ and ${\bar v}_\chi$ are the local neutralino density
and rms velocity in the halo respectively, $\sigma_i$ is the
elastic scattering cross-section of the neutralino with the
nucleus of element $i$ with mass $m_i$, $f_i$ is the mass fraction
of element $i$, and $X_i$ is a kinematic factor which accounts
for several important effects: (1) the motion of the Sun or
Earth relative to Galactic center; (2) the suppression due to the
mismatching of $m_\chi$ and $m_i$; (3) the loss of coherence in the
interaction due to the finite size of the nucleus
(see Ref.~\cite{Gould} for details).

In the summation in Eq.~(\ref{eq:nc}), we only include the ten most abundant
elements for the Sun or Earth respectively, and use the mass
fraction $f_i$ of these elements as listed in Table A.1
of Ref.~\cite{GGR91}. We choose
${\bar v}_\chi = 300\,{\rm km}\,{\rm sec}^{-1}$, a value within
the allowed range of the characteristic velocity of halo dark matter
particles. To take into account the effect of the actual
neutralino relic density, we follow the conservative
approach of Ref.~\cite{GGR91} for the local neutralino density $\rho_\chi$:
(a) $\rho_\chi = \rho_h = 0.3\,{\rm GeV}/{\rm cm}^3$,
if $\Omega_\chi h^2_0 > 0.05$;
while (b) $\rho_\chi = (\Omega_\chi h^2_0/0.05)\rho_h$,
if $\Omega_\chi h^2_0 \lsim 0.05$.
As for $\sigma_i$, the dominant contribution is the coherent
interaction due to the exchange of two CP-even Higgs bosons $h$ and $H$
and squarks, and we use the expressions (A10) and (A11)
of Ref.~\cite{KAM}\footnote{We have corrected a sign error for
the $H\chi\chi$ coupling in (A10) of Ref.~\cite{KAM}.}
to compute the spin-independent cross section for all the elements
included. In addition, for capture by the Sun, we also evaluate
the spin-dependent cross section due to both $Z$-boson exchange
and squark exchange for the scattering from hydrogen according
to Eq.~(A5) (EMC model case) of Ref.~\cite{KAM}. It should be noted that in all
these expressions the squarks were assumed to be degenerate. In the two
supergravity models that we consider here this need
not be the case, although for most of the parameter space this
is a fairly good approximation. Hence, we simply
use the average squark mass $m_{\tilde q}$ in this part of the
calculation.

The kinematic factor $X_i$ in Eq.~(\ref{eq:nc}), can be most
accurately evaluated once the detailed knowledge of the mass
density profile as well as the local escape velocity profile
are specified for all the elements.
In practice, this can be done by performing a numerical integration
with the physical inputs provided by standard solar model or some
sort of Earth model. Instead of performing such an involved calculation,
we approximate the integral for each element by the value of the
integrand obtained with the average effective gravitational
``potential energy'' $\phi_i$ times the integral volume. The
values of $\phi_i$ are taken from Table A.1 of Ref.~\cite{GGR91}.

\section {The Detection Rate}
\label{sec:det}

We next describe the procedure employed by us to calculate the
detection rate of upwardly-moving muons, resulting from the
particle production and interaction subsequent to the capture
and annihilation of neutralinos in the two supergravity models.
The annihilation process normally reaches equilibrium with the capture
process on a time scale much shorter than the age of the Sun or Earth.
We assume this is the case, so that the neutralino annihilation rate
equals half of the capture rate.
The detection rate for neutrino-induced
upwardly-moving muon events is then given by
\begin{equation}
\Gamma={C\over 8\pi R^2}\sum_{i,F}{D_iB_F\int
\left({dN\over dE_\nu}\right)_{iF}{E^2_\nu}dE_{\nu}}.\label{eq:nd}
\end{equation}
In Eq.~(\ref{eq:nd}), $D_i$ is a constant, $R$ is the distance
between the detector and the Sun or the center of the Earth,
and $(dN/dE_\nu)_{iF}$ is the differential energy spectrum of
neutrino type $i$ as it emerges at the surface of the Sun or Earth
due to the annihilation of neutralinos in the core of the Sun or Earth
into final state $F$ with a branching ratio $B_F$. It should be noted
that in Eq.~(\ref{eq:nd}) that $i$ is summed over muon neutrinos and
anti-neutrinos, and that $F$ is summed over final states that contribute
to the high-energy neutrinos. The only relevant fermion pair
final states are $\tau{\bar\tau}$, $c{\bar c}$, $b{\bar b}$, and (for
the $SU(5)\times U(1)$ model) $t{\bar t}$ when $m_\chi > m_t$. The
lighter fermions do not produce high-energy neutrinos since
they are stopped by the solar or terrestrial media before
they can decay \cite{RS}.

The branching ratio $B_F$ can be easily calculated as the relative magnitude
of the thermal-averaged product of annihilation cross section into final
state $F$ ($\sigma_F$) with the M\o ller velocity $v_M$.
Since the core temperatures of
the Sun and Earth are very low compared with the neutralino mass
($T_{\rm Sun} \sim 1.34\times 10^{-6}$ GeV;
$T_{\rm Earth} \sim 4.31\times 10^{-10}$ GeV), only the $s$-wave
contributions are relevant, hence it is enough here to use the
usual thermal average expansion up to zero-order of $T/m_\chi$
($v_M \rightarrow 0$ limit), {\it i.e.}
\begin{equation}
B_F={\langle{\sigma_F v_M}\rangle\over
\langle{\sigma_{\rm tot} v_M}\rangle}={a_F\over a_{\rm tot}}.
\label{eq:bratio}
\end{equation}
In Eq.~(\ref{eq:bratio}), {\it all} kinematically allowed final
states contribute to $a_{\rm tot}$. Besides all the fermion
pair final states, we have also included boson pair final states
$WW$, $ZZ$ and $hA$ in our calculation of $B_F$.
Due to the parameter space constraints, these channels are not
open for the minimal $SU(5)$ model. But $WW$ and $ZZ$ channels
are generally open in the $SU(5)\times U(1)$ model, and the $hA$ channel
also opens up for large values of $\tan\beta$ in the
dilaton case. We should also remark that the annihilation channel
into lightest CP-even higgs pair $hh$ is always allowed kinematically
in some portion of the parameter space for both supergravity models
we consider, but since its $s$-wave contribution vanishes,
we do not include it in the calculation of $B_F$. However,
this channel is taken into account in the calculation of the
neutralino relic density, which does affect the capture
rate through the scaling of local density $\rho_\chi$
when $\Omega_\chi h^2_0 < 0.05$ (see Sec.~\ref{sec:cap}).
In addition, we have kept all the nonvanishing
interference terms in the evaluation of $a_{\rm tot}$ and $a_F$.

The calculation of the neutrino differential energy
spectrum is somewhat involved, since it requires a reasonably accurate
tracking of the cascade of the particles which result from neutralino
annihilation into each of the final state $F$. This involves the decay
and hadronization of the various annihilation products and their
interactions with the media of the Sun or the Earth's cores.
In addition, at high energies, neutrinos interact
with and are absorbed by  solar matter, a fact that affects the spectrum.
In Ref.~\cite{RS}, Ritz and Seckel rendered this calculation tractable
by their adaptation of the Lund Monte Carlo for this purpose. Subsequently,
analytic approximations to the Monte Carlo procedure outlined in
their paper were refined and employed by Kamionkowski to calculate
the neutrino energy spectra from neutralino annihilation for
the MSSM in Ref.~\cite{KAM}. The procedure involved is described
below for completeness.

Since the probability for producing an underground muon
that traverses the detector is proportional to the square of
the neutrino energy, the  primary quantity of interest is
the second moment $\langle{Nz^2}\rangle m^2_\chi$, defined as
\begin{equation}
{\langle Nz^2 \rangle}_{iF} \equiv{1\over m^2_\chi}\int
\left({dN\over dE_{\nu}}\right)_{iF}{E^2_\nu}dE_\nu.
\label{eq:secmom}
\end{equation}
Once the second moments are obtained, the detection rate
for the neutrino-induced upwardly-moving muon events (\ref{eq:nd})
may be conveniently written as
\begin{equation}
\Gamma=\kappa_B ({C\over {\rm sec}^{-1}}) ({m_\chi\over {\rm GeV}})^2
\sum_i{a_ib_i}\sum_F{B_F {\langle Nz^2 \rangle}_{iF}}\ \
{\rm m}^{-2}\,{\rm yr}^{-1},
\label{eq:drt}
\end{equation}
where $\kappa_B =1.27\times 10^{-29}$ ($7.11\times 10^{-21}$) for
neutrinos from the Sun (Earth), the scattering coefficients
$a_i=6.8$ ($3.1$) for neutrinos (anti-neutrinos),
while the muon range constants $b_i=0.51$ ($0.67$) for neutrinos
(anti-neutrinos).

The approximations of Refs.~\cite{RS,KAM} consist in obtaining expressions
for the second moment without detailed knowledge of the functional
form of the differential energy spectra. We now list these approximate
expressions for ${\langle Nz^2 \rangle}_{iF}$ as follows:

(i) {\it Fermion pair final states}
($\tau{\bar\tau}, c{\bar c}, b{\bar b}, t{\bar t}$).
The simplest case is that of fermions injected into the core of the
Earth, since all interactions are negligible. We use
\begin{equation}
\langle Nz^2 \rangle={1\over 3}
\langle N \rangle \langle y^2 \rangle
\left[\langle z^2_f \rangle -{m^2_f\over 4E^2_{in}}\right].
\label{eq:fma}
\end{equation}
Here $\langle N \rangle$ and $\langle y^2 \rangle$
are the rest frame yield and second moments, while
$\langle z^2_f \rangle$ is the second moment of the fragmentation
function, obtained from Table 2 and Table 3 respectively in \cite{RS}.
$m_f$ and $E_{in}$ are the mass and energy of the injected fermion.

Fermions injected into the core of the Sun undergo interactions and decay
before final state neutrinos emerge at the surface. These are analytically
approximated reliably for fermion injection energies relevant to our
situation by \cite{KAM}
\begin{equation}
\langle Nz^2 \rangle
=\left[ax_0e^{x_0}\int_{x_0}^{\infty}\frac{e^{-x}}{x}dx\right]^2,
\label{eq:fmb}
\end{equation}
where $x_0=155/E_{in}$ ($275/E_{in}$) and $a=0.056$ ($0.052$)
for neutrinos (anti-neutrinos) from $c$ quarks,
while $x_0=185/E_{in}$ ($275/E_{in}$) and $a=0.086$ ($0.082$)
for neutrinos (anti-neutrinos) from $b$ quarks.
For the top quark we have considered three different
masses: 130, 150 and 170 GeV in the $SU(5)\times U(1)$ model.
In this case the approximations are less reliable \cite{KAM},
but better at energies below the $100$ GeV scale than above it.
The appropriate values here are
$a=0.18$ ($0.14$) and $x_0=110/E_{in}$ ($380/E_{in}$)
for neutrinos (anti-neutrinos).
In the above (and in what follows), $E_{in}$ and other energies
are taken in GeV when obtaining numerical values.
The above expressions include the hadronization of $b$ and $c$ and $t$ quarks
in the solar medium. The $\tau$ lepton decays almost instantly, and
does not hadronize, so must be treated differently. The second moment
for the $\tau$ is approximated by
\begin{equation}
\langle Nz^2 \rangle =ae^{-E_{in}/E_a}
\label{eq:fmc}
\end{equation}
where $a=0.0204$ ($0.0223$) and $E_a=476$ ($599$) for neutrino
(anti-neutrino) production.

(ii) {\it Gauge boson pair final states} ($WW$, $ZZ$).
The production of $\nu_\mu$ and ${\bar \nu}_\mu$ by $W,Z$ boson pair
and Higgs boson pair is related to that by fermions, since the  bosons
primarily decay into fermions, hence the subsequent cascade
and hadronization remains the same.
For the annihilation final state $WW$, the $W$ boson with velocity
$\beta$ and energy $E$ will decay into $\mu{\nu_\mu}$ with
a fractional width $\Gamma(W\rightarrow \mu {\bar \nu}_\mu)$. The second moment
for this annihilation final state for the neutrinos from the Sun
is \cite{KAM}
\begin{equation}
\langle Nz^2 \rangle
=\left[{\Gamma(W\rightarrow \mu {\bar \nu}_\mu)\over \beta E^3}
{2+2E_{in}\tau_i(1+\alpha_i)+E_{in}^{2}\tau^2_i\alpha_i(1+\alpha_i)
\over \alpha_{i}\tau^3_i(\alpha^2_i-1)(1+E_{in}\tau_i)^{1+\alpha_i}}
\right]^{E_{in}=E(1-\beta)/2}
_{E_{in}=E(1+\beta)/2}.
\end{equation}
Here $\tau_{i}=1.01 \times 10^{-3}$ GeV$^{-1}$
($3.8 \times 10^{-4}$ GeV$^{-1}$),
$\alpha_i =5.1$ ($9.0$) for neutrinos (anti-neutrinos).
For the neutrinos from the Earth, the second moments is approximated
by a simpler expression
\begin{equation}
\langle Nz^2 \rangle = \Gamma(W\rightarrow \mu {\bar \nu}_\mu)(3+\beta^2)/12
\end{equation}
due to the absence of interactions. The expressions for the $ZZ$
final state can be similarly obtained \cite{KAM}.

(iii) {\it Higgs boson pair final state} ($hA$).
For the annihilation final state $hA$,
each Higgs boson $S$ ($=h,A$) with velocity $\beta_S$ and energy $E_S$
will decay into fermion pair $f{\bar f}$ with a branching ratio
$BR(S\rightarrow f{\bar f})$. Assuming such decays are isotropic,
the second moment of this final state is
\begin{equation}
{\langle Nz^2 \rangle}
=\sum_{S=h,A}\sum_f
{2BR(S\rightarrow f{\bar f})\over \beta_S E_S}
\int^{E_S(1+\beta_S)/2}_{E_S(1-\beta_S)/2}
{\langle Nz^2 \rangle}_fdE_f.
\label{eq:hA}
\end{equation}
Here $E_f$ is the fermion energy, and ${\langle Nz^2 \rangle}_f$ is the
second moment for the fermion pair $f{\bar f}$
($\tau{\bar\tau}, c{\bar c}, b{\bar b}, t{\bar t}$) as given in
Eqs.~(\ref{eq:fma})--(\ref{eq:fmc}).

\section{Results and Discussion}
\label{sec:result}
For each point in the parameter spaces of the two models described
in Sec.~\ref{sec:model}, namely the minimal $SU(5)$ and the
$SU(5)\times U(1)$ supergravity models (both the no-scale and dilaton
cases), we have determined the relic abundance of neutralinos and
then computed the capture rate in the Sun and Earth
(as described in Sec.~\ref{sec:cap}) and
the resulting upwardly-moving muon detection rate
(as described in Sec.~\ref{sec:det}).
In Figs.~\ref{NT1} and \ref{NT2}, the predicted capture and detection rates
in the minimal $SU(5)$ supergravity model are shown,
based on the assumption that the mass of the triplet higgsino,
which mediates dimension-five proton decay, obeys $M_{{\tilde H}_3}<3M_U$.
We have redone the calculation relaxing this assumption to
$M_{{\tilde H}_3}<10M_U$, in which case the results for the muon fluxes
remain qualitatively the same, except that the parameter space is opened
up somewhat. The dashed lines in Fig.~\ref{NT2} represent the current
Kamiokande 90\% C.L. upper limits (1) and (2).
Similarly, the predictions of the $SU(5)\times U(1)$
supergravity model are presented in Figs.~\ref{NT3} and \ref{NT4}
for the no-scale scenario, and in Figs.~\ref{NT5} and \ref{NT6} for
the dilaton scenario, again along with the Kamiokande upper limits (dashed
lines). In Figs.~\ref{NT3}--\ref{NT6}, we have taken the representative value
of $m_t=150$~GeV. Similar results are obtained for other values of $m_t$.

Several comments on these figures are in order. First, the kinematic
enhancement of the capture rate by the Earth manifests itself
in all figures as the big peaks near the Fe mass ($m_{\rm Fe}=52.0$ GeV),
as well as the smaller peaks around Si mass ($m_{\rm Si}=26.2$ GeV).
Second, there is a severe depletion of the rates near
$m_\chi={1\over 2} M_Z$ in Figs.~\ref{NT3}--\ref{NT6}, which is
due to the decrease in the neutralino relic density. In the case of
Earth capture, this effect is largely compensated by the
enhancement near the Fe mass. As mentioned in Sec.~\ref{sec:cap},
in our procedure, the relic density affects the local
neutralino density $\rho_\chi$ only if $\Omega_\chi h^2_0 < 0.05$,
while in the minimal $SU(5)$ model this almost never happens,
therefore, the effect of the $Z$-pole is not very evident in Figs.~\ref{NT1}
and \ref{NT2}. Also, in Figs.~\ref{NT3}--\ref{NT6}, the various
dotted curves correspond to different values of $\tan\beta$,
starting from the bottom curve with $\tan\beta = 2$, and increasing
in steps of two. These curves clearly show that the capture and
detection rates increase with increasing $\tan\beta$, since
the dominant piece of the coherent neutralino-nucleon scattering
cross section via the exchange of the lightest Higgs boson $h$ is
proportional to $(1+{\tan}^2\beta)$. The capture rate decreases
with increasing $m_\chi$, since the scattering cross section
falls off as $m^{-4}_h$ and $m_h$ increases with increasing $m_\chi$.
It is expected that the detection rate in general also decreases
for large value of $m_\chi$, since it is proportional to the capture
rate. However, the opening of new annihilation channels,
such as the $WW$, $ZZ$ and $hA$ channels in the $SU(5)\times U(1)$ model,
could have two compensating effects on the detection rate: (a) the
presence of a new channel to produce high-energy neutrinos
which leads to an enhancement of the detection rate; and (b) the
decrease of the branching ratios for the fermion pair channels,
which makes the neutrino yield from $\tau{\bar\tau}$,
$c{\bar c}$ and $b{\bar b}$ smaller and hence reduces the
detection rate. Therefore, these new annihilation channels
could {\it either} increase {\it or} decrease the detection
rate, depending which of these two effects wins over.
We found that, for small values of $\tan\beta$ and $\mu < 0$,
the $WW$ channel can become dominant if open,
basically because in this case the neutralino contains a rather
large neutral wino component. This explains the distortion of
the detection rate curves in the $\mu < 0$ half of Figs.~\ref{NT4}
and \ref{NT6}. The effect of the $ZZ$ channel turns out to be
negligible in the $SU(5)\times U(1)$ model, since neutralinos
with $m_\chi > M_Z$ have very small higgsino components. The same
argument applies to the $hA$ channel which sometimes opens up
in the dilaton case for rather large values of $\tan\beta$.

In the dilaton case, for large values of $\tan\beta$
the CP-odd Higgs boson $A$ can be rather light, and
the presence of the $A$-pole when $m_\chi \sim {1\over 2}m_A$
makes the relic density very small. $\Omega_\chi h^2_0$ as a function
of $m_\chi$, is first lower than $0.05$, it increases with $m_\chi$,
and eventually reaches values above $0.05$, when neutralinos move away
from the $A$-pole. Thus, the capture and detection rates also show
this behavior, which can be seen as the few ``anomalous'' lines in
Figs.~\ref{NT5} and \ref{NT6}. For lower values of $\tan\beta$,
$\Omega_\chi h^2_0 < 0.05$, and there is no such effect.
In the minimal $SU(5)$ model, since the allowed points include
different supersymmetry breaking scenarios and several values
of $m_t$ and $\tan\beta$, all these features are blurred.
Nonetheless, in the same range of $m_\chi$ and for same
values of $\tan\beta$ and $m_t$, we have found the results
of these two models comparable, with the rates in the
$SU(5)\times U(1)$ model slightly smaller due to the smaller
relic density.

It is clear that at present the experimental constraints
from the ``neutrino telescopes'' on the parameter space
of the two supergravity models are quite weak.
In fact, only the Kamiokande upper bound from the Earth
can be used to exclude regions of the parameter
space with $m_\chi \approx m_{\rm Fe}$ for both models,
in particular for the $SU(5)\times U(1)$ model, due to
the enhancement effect discussed above. However, it is our belief
that the results presented in this paper will be quite useful in
the future, when improved sensitivity in underground muon detection
rates become available. An improvement in experimental sensitivity
 by a factor of two should be easily possible when MACRO \cite{mac} goes
into operation, while a ten-fold improvement is envisaged when
Super-Kamiokande \cite{sk} announces its results sometime by the
end of the decade. More dramatic improvements in the sensitivity
(by a factor of 20--100) may be expected from DUMAND and
AMANDA \cite{amd}, currently under construction.
In addition, as recently argued \cite{hal},
we think that perhaps the full parameter space of a large class
of supergravity models, including the two specific ones considered here,
may only be convincingly probed by a detector with an effective
area of 1 km$^2$. It is interesting to note that, with a sensitivity
improvement by a factor of 100, a large portion of the $\mu < 0$
half parameter space of the minimal $SU(5)$ model can be probed.
Unfortunately, the remaining portion, with fluxes below $\sim10^{-4}$,
can hardly be explored by underground experiments in the foreseeable future.
For the $SU(5)\times U(1)$ supergravity models, in Figs.~\ref{NT7}
and \ref{NT9} we have plotted the allowed points in the
$(m_{\chi^\pm_1},\tan\beta)$ space. These points are those obtained originally
in Refs.~\cite{LNZI,LNZII}, such that the neutrino telescope constraint is
also satisfied; no other constraints have been imposed. It can be seen here
that the small voids of points for $m_{\chi^\pm_1}\approx100$~GeV and a variety
of values of $\tan\beta$ are excluded by the constraint from the ``neutrino
telescopes". In these figures we have marked by crosses
the points in the $(m_{\chi^\pm_1},\tan\beta)$ plane that
MACRO should be able to probe (assuming an increase in the sensitivity
by a factor of two) for the no-scale and dilaton
scenarios respectively. Finally, Figs.~\ref{NT8} and \ref{NT10}
show that, using this indirect technique, Super-Kamiokande can
cover nearly half of the parameter space of the $SU(5)\times U(1)$
model, assuming that an improvement by a factor of about twelve can
be achieved. As expected, the  constraints from future
``neutrino telescopes'' will be strictest for large values
of $\tan\beta$.

To keep the above results in perspective, we now discuss
some effects that have not been taken into account in our analysis.
As we have shown, for the two supergravity models, the capture
of the neutralinos by the Earth is more important than by
the Sun. In the Earth case, we have considered only the
{\it primary direct} capture of the neutralinos from the
galactic halo, in which a neutralino is trapped
by the Earth's gravitational field only when its velocity falls
below the escape velocity at a point inside the Earth,
as a consequence of its interaction with the nuclei around.
However, there exists yet another mechanism by which neutralinos
can be captured by the Earth, namely, the {\it secondary indirect}
capture, first studied by Gould \cite{Gould}. In this mechanism,
a neutralino first loses only enough energy
to be bound in a solar orbit (``orbit capture''), but not
enough to be directly captured by the Earth. The orbit-captured neutralinos
further weakly interact with the nuclei in the Earth, and a fraction
of them subsequently can be indirectly captured into the Earth's
core. It was found by Gould that, for Dirac neutrinos of mass
10--80 GeV, direct and indirect capture by the Earth are of the
same order of magnitude, and the kinematic enhancement of the
total capture rate becomes enlarged and broadened \cite{Gould}.
We believe that this result should also apply to the case of neutralino
capture, which means that the Earth capture rate that
we have computed in this work may have been significantly {\it underestimated}.
Although to our knowledge the indirect capture was also not
considered in recent works that dealt with Earth
capture \cite{GGR91,KEK,Bott,FH}, we feel that this important
mechanism should be taken into account in future analyses of
this type. In fact, from Figs.~\ref{NT7} and \ref{NT9}, we see
that even before MACRO, interesting constraints may already be
extracted from the current Kamiokande upper limits, had we
included the indirect capture in our calculation.

Recently, after the calculations in this paper had been completed,
we received Refs.~\cite{DNI,DNII} in which some other issues
of particle physics relevant to our work have been addressed.
Given the fact that the neutralino in the two supergravity models we study
is mostly a ``mixed'' state, in particular for
$m_\chi \sim m_{\rm Fe}$, according to Fig.~7 of Ref.~\cite{DNI},
we expect that the results of Ref.~\cite{DNI}, when they become appreciable,
could lead to a shift of our detection rates in both directions by
about $10\%$ or less. We have not redone our
calculations of the capture rate using the cross section of
Ref.~\cite{DNII}. The inclusion of the new
neutralino-gluon scatterings \cite{DNII}, when they become appreciable,
would increase the chance that neutralinos can be captured
by {\it both} the Sun {\it and} the Earth,
barring the normally small interference effects which sometimes
could render the total cross section of Ref.~\cite{DNII} smaller
than what it would be without these new scatterings \cite{Nojiri}.
Therefore, the effect of the results of Ref.~\cite{DNII} on
the detection rate, could somehow enhance or compensate
that of Ref.~\cite{DNI}, depending whether the latter
is an increase or decrease of the detection rate.
However, even as a conservative estimate, we do not expect that
the combined effect of Refs.~\cite{DNI,DNII} would alter our
results quantitatively by more than $20\%$, a change not large enough to affect
our conclusions qualitatively.

We have also looked into the possibility that the high energy neutrino
flux may be altered by MSW oscillations \cite{msw} in the Sun. This issue
has been addressed in a recent paper \cite{erfm}. From their results we
conclude that for the neutrino energy range relevant to this work,
$\nu_{\mu}$ to $\nu_{e}$ oscillations are negligible. The
$\nu_{\mu}$ flux may also be altered by $\nu_{\mu}$ to $\nu_{\tau}$
vacuum oscillations during passage from the Sun to the Earth. This effect
will be significant only if the $\nu_{\mu}$-$\nu_{\tau}$ mixing angle is
large. Large mixing is disfavored by the general GUT based see-saw arguments
\cite{ssaw}. An analysis based on these considerations coupled with
phenomenological arguments in the context of the $SU(5)\times U(1)$
model \cite{eln} supports a value for $\sin^{2}2\theta\approx 10^{-4}$,
which is much too small to affect the flux values predicted here.

\section{Conclusions}
\label{sec:conclusion}

We have explored the possibility of detecting supersymmetry indirectly through
the measurement of upwardly-moving muons in underground detectors or
``neutrino telescopes''. These muons originate from high-energy neutrino
interactions in the rock below the detector, and these neutrinos result from
annihilation of neutralinos in the Earth or Sun cores. The latter would have
been captured by these heavenly bodies if the neutralinos constitute an
important part of the galactic halo---an important assumption which should
not be overlooked.

The present day experimental upper bounds on the muon flux are only weakly
constraining. In fact, this is mostly because the large possible fluxes for
light neutralinos (see for example Fig.~\ref{NT4} for small $m_\chi$) have
already been ruled out by the LEP lower bounds on the neutralino mass.
Nonetheless, there is a region of parameter space with
$m_{\chi^\pm_1}\approx100$~GeV (corresponding to $m_\chi\approx m_{\rm Fe}$)
and a range of values of $\tan\beta$, which is excluded at the 90\% C.L.
However, expected increases in experimental sensitivity in the next few years
should turn this technique into a very efficient way of probing the parameter
space of the specific supergravity models considered here, even surpassing
the reach of traditional direct detection collider experiments.

We should remark that even though our explicit computations apply only
to the $SU(5)$ and $SU(5)\times U(1)$ models, the correlations among sparticle
and Higgs boson masses, which play such a fundamental role in the quantitative
results, are common to a large class of supergravity models with radiative
electroweak symmetry breaking \cite{aspects}. In this respect, the specific
models considered here include values of $\xi_0=m_0/m_{1/2}$ which are small
($\xi_0=0$, no-scale), moderate ($\xi_0=1/\sqrt{3}$, dilaton), and large
($\xi_0\gg1$, minimal $SU(5)$), and therefore span the whole allowed range.
Thus, our results: (i) represent a significant sampling of what would be
obtained using an arbitrary selection of parameters in a generic supergravity
model, and (ii) will allow to select the correct supergravity model when
experimental data start to restrict the parameter space.

\section*{Acknowledgments}

This work has been supported in part by DOE grant DE-FG05-91-ER-40633.
The work of R.G. and K.Y. has been supported by a World-Laboratory Fellowship.
The work of J.L. has been supported by an SSC Fellowship.

\vfill\eject

\vfill\eject
\begin{figure}
\noindent {\Large\bf Figure Captions}\\
\caption{The neutralino capture rate for the Sun and Earth as a function of
the neutralino mass in the minimal $SU(5)$ supergravity model.}
\label{NT1}
\caption{The upwardly-moving muon flux in underground detectors originating
from neutralino annihilation in the Sun and Earth, as a function of the
neutralino mass in the minimal $SU(5)$ supergravity model. The dashed
lines represent the present Kamiokande 90\% C.L. experimental upper limits.}
\label{NT2}
\caption{The neutralino capture rate for the Sun and Earth as a function of the
neutralino mass in the no-scale $SU(5)\times U(1)$ supergravity model. The
representative value of $m_t=150$~GeV has been used. Note the depletion of
neutralinos in the halo near the $Z$-resonance, and the enhancement in the
Earth capture rate near the iron nucleus mass (52.0 GeV).}
\label{NT3}
\caption{The upwardly-moving muon flux in underground detectors originating
from neutralino annihilation in the Sun and Earth, as a function of the
neutralino mass in the no-scale $SU(5)\times U(1)$ supergravity model. The
representative value of $m_t=150$~GeV has been used. The dashed lines
represent the present Kamiokande 90\% C.L. experimental upper limits.}
\label{NT4}
\caption{Same as Fig.~3 but for the dilaton $SU(5)\times U(1)$ supergravity
model.}
\label{NT5}
\caption{Same as Fig.~4 but for the dilaton $SU(5)\times U(1)$ supergravity
model.}
\label{NT6}
\caption{The allowed parameter space of the no-scale $SU(5)\times U(1)$
supergravity model (in the $(m_{\chi^\pm_1},\tan\beta$) plane) after the
present ``neutrino telescopes" (NT) constraint has been applied. Two values
of $m_t$ (130,150 GeV) have been chosen. The crosses denote those points which
could be probed with an increase in sensitivity by a factor of two.}
\label{NT7}
\caption{Same as Fig.~7 but the crosses now denote points which could be
probed with an increase in sensitivity by a factor of 12.}
\label{NT8}
\caption{Same as Fig.~7 but for the dilaton $SU(5)\times U(1)$ supergravity
model.}
\label{NT9}
\caption{Same as Fig.~8 but for the dilaton $SU(5)\times U(1)$ supergravity
model.}
\label{NT10}
\end{figure}


\begin{thebibliography}{99}
\bibitem{Haber}See {\em e.g.}, H. Haber, SCIPP-93/22, to appear in
Proceedings of the HARC Workshop on Recent Advances in the Superworld,
The Woodlands, April 1993.
\bibitem{ewcorr} J.L. Lopez, D.V. Nanopoulos, G.T. Park, H. Pois, and
K. Yuan, Phys. Rev. D {\bf48} (1993) 3297.
\bibitem{bsgamma} J.L. Lopez, D.V. Nanopoulos, and G.T. Park,
Phys. Rev. D {\bf48} (1993) R974;
J.L. Lopez, D.V. Nanopoulos, G.T. Park, and A.Zichichi,
Texas A\&M preprint CTP-TAMU-40/93.
\bibitem{g-2} J.L. Lopez, D.V. Nanopoulos, and X. Wang, Texas A\&M
preprint CTP-TAMU-44/93.
\bibitem{PS85}W.H. Press and D.N. Spergel, Astrophys. J. {\bf 296}
(1985) 679.
\bibitem{Gould}A. Gould,  Astrophys. J. {\bf 321} (1987) 560, 571;
{\bf 328} (1988) 919; {\bf 388} (1992) 338.
\bibitem{others} J. Silk, K. Olive, and M. Srednicki, Phys. Rev. Lett.
{\bf 55} (1985) 257; T. Gaisser, G. Steigman, and S. Tilav, Phys. Rev.
D {\bf34} (1986) 2206; J. Hagelin, K. Ng, and K.  Olive,
Phys. Lett. B {\bf180} (1987) 375; M. Srednicki, K. Olive, and J. Silk,
Nucl. Phys. B {\bf 279} (1987) 804; K. Ng, K. Olive, and M. Srednicki,
Phys. Lett. B {\bf188} (1987) 138;
K. Olive and M. Srednicki, Phys. Lett. B {\bf205} (1988) 553; L. Krauss,
M. Srednicki, and F. Wilczek, Phys. Rev. D {\bf33} (1986) 2079; K. Freese,
Phys. Lett. {\bf 167B} (1986) 295.
\bibitem{RS}S. Ritz and D. Seckel, Nucl. Phys. B {\bf 304} (1988) 877.
\bibitem{GR89}G.F. Giudice and E. Roulet, Nucl. Phys. B {\bf 316} (1989)
429.
\bibitem{GGR91}G. Gelmini, P. Gondolo, and E. Roulet, Nucl. Phys.
B {\bf 351} (1991) 623.
\bibitem{KAM}M. Kamionkowski, Phys. Rev D {\bf 44} (1991) 3021.
\bibitem{KEK}M. Mori {\it et al.} (Kamiokande Collaboration),
Phys. Lett. B {\bf 270} (1991) 89.
\bibitem{Bott}A. Bottino, V. de Alfaro, N. Fornengo, G. Mignola,
and S. Scopel, Phys. Lett. B {\bf 265} (1991) 57;
Astroparticle Phys. {\bf 1} (1992) 61.
\bibitem{FH}F. Halzen, M. Kamionkowski, and T. Stelzer, Phys. Rev D
{\bf 45} (1992) 4439.
\bibitem{aspects}See {\em e.g.}, S. Kelley, J.L. Lopez, D.V. Nanopoulos,
H. Pois, and K. Yuan, Nucl. Phys. B {\bf 398} (1993) 3.
\bibitem{KEKII}M. Mori {\it et al.} (Kamiokande Collaboration),
Phys. Lett. B {\bf 289} (1992) 463.
\bibitem{mac}B. Barish in the Proceedings of the Supernova Watch Workshop
(Santa Monica, California, 1990) (unpublished).
\bibitem{sk}Y. Suzuki in Proceedings of the 3rd International Workshop
on Neutrino Telescopes, Venice, March 1992, ed. M. Baldo-Ceolin, Venice
(1992).
\bibitem{amd}J. Learned in Proceedings of the European Cosmic Ray
Symposium, Geneva, July 1992, eds. P. Grieder and B. Pattison,
Nucl. Phys. B (1993), to appear.
\bibitem{Dickreview}For reviews see {\em e.g.}, R. Arnowitt and P. Nath,
{\it Applied N=1 Supergravity} (World Scientific, Singapore 1983);
H. P. Nilles, Phys. Rep. {\bf110} (1984) 1;
J.L. Lopez, D.V. Nanopoulos, and A. Zichichi, CERN-TH.6934/93 and
Texas A\&M University preprint CTP-TAMU-34/93.
\bibitem{EriceDec92} For a recent review see J.L. Lopez, D.V. Nanopoulos,
and A. Zichichi, CERN-TH.6926/93 and Texas A\&M University
preprint CTP-TAMU-33/93.
\bibitem{EKN}J. Ellis, S. Kelley, and D.V. Nanopoulos, Phys. Lett.
B {\bf249} (1990) 441, B {\bf260} (1991) 131, B {\bf 287} (1992) 95,
Nucl. Phys. B {\bf 373} (1992) 55; P. Langacker and M.-X. Luo, Phys. Rev.
D {\bf44} (1991) 817;
F. Anselmo, L. Cifarelli, A. Peterman,
and A. Zichichi, Nuovo Cim. {\bf104A} (1991) 1817 and {\bf105A} (1992)
581,1025,1179,1201; F. Anselmo, L. Cifarelli, and A. Zichichi,
Nuovo Cim. {\bf105A} (1992) 1335,1357;
G. Ross and R. Roberts, Nucl. Phys. B {\bf 377} (1992) 571;
P. Langacker and N. Polonsky, Phys. Rev. D {\bf47} (1993) 4028;
R. Barbieri and L. Hall, Phys. Rev. Lett. {\bf68} (1992) 752;
L. Hall and U. Sarid, Phys. Rev. Lett. {\bf70} (1993) 2673.
\bibitem{ANpd}R. Arnowitt and P. Nath, Phys. Rev. Lett. {\bf69} (1992) 725;
P. Nath and R. Arnowitt, Phys. Lett. B {\bf287} (1992) 89.
\bibitem{susydm}J.L. Lopez, K. Yuan, and D.V. Nanopoulos,
Phys. Lett. B {\bf267} (1991) 219;
S. Kelley, J.L. Lopez, D.V. Nanopoulos, H. Pois, and K. Yuan,
Phys. Rev. D {\bf47} (1993) 2461.
\bibitem{troubles}J.L. Lopez, D.V. Nanopoulos, and A. Zichichi,
Phys. Lett. B {\bf291} (1992) 255; J.L. Lopez, D.V. Nanopoulos, and H. Pois,
Phys. Rev. D {\bf47} (1993) 2468; R. Arnowitt and P. Nath,
Phys. Lett. B {\bf299} (1993) 58 and {\bf307} (1993) 403(E);
P. Nath and R. Arnowitt, Phys. Rev. Lett. {\bf70} (1993) 3696;
J.L. Lopez, D.V. Nanopoulos, and K. Yuan, Phys. Rev. D {\bf48} (1993) 2766.
\bibitem{LNPZ}J.L. Lopez, D.V. Nanopoulos, H. Pois, and A. Zichichi,
Phys. Lett. B {\bf299} (1993) 262.
\bibitem{Lacaze}I. Antoniadis, J. Ellis, R. Lacaze, and D.V. Nanopoulos,
Phys. Lett. B {\bf268} (1991) 188; S. Kalara, J.L. Lopez, and D.V. Nanopoulos,
Phys. Lett. B {\bf269} (1991) 84.
\bibitem{Search}I. Antoniadis, J. Ellis, S. Kelley, and D.V. Nanopoulos,
Phys. Lett. B {\bf272} (1991) 31; J.L. Lopez, D.V. Nanopoulos, and K. Yuan,
Nucl. Phys. B {\bf 399} (1993) 654.
\bibitem{LNZI}J.L. Lopez, D.V. Nanopoulos, and A. Zichichi, CERN-TH.6667/92,
Texas A\&M University preprint CTP-TAMU-68/92.
\bibitem{LN}For a review see A. B. Lahanas and D.V. Nanopoulos, Phys. Rep.
{\bf 145} (1987) 1.
\bibitem{LNZII}J.L. Lopez, D.V. Nanopoulos, and A. Zichichi, CERN-TH.6903/93,
Texas A\&M University preprint CTP-TAMU-31/93.
\bibitem{dilaton}V. Kaplunovsky and J. Louis, Phys. Lett.
B {\bf 306} (1993) 269; R. Barbieri, J. Louis, and M. Moretti,
Phys. Lett. B {\bf 312} (1993) 451.
\bibitem{ADD1}See {\it e.g.}, L. Roszkowski, Phys. Lett. B {\bf262} (1991) 59.
\bibitem{ADD2}See {\it e.g.}, J.L. Lopez, D.V. Nanopoulos, and K. Yuan,
Nucl. Phys. B {\bf 370} (1992) 445.
\bibitem{hal} F. Halzen, Talk presented at the Johns Hopkins Workshop
on particles and the Universe, Budapest, July 1993, MAD-PH-785 (1993).
\bibitem{DNI}M. Drees, G. Jungman, M. Kamionkowski, and M.M. Nojiri,
EFI-93-38, MAD-PH-766 (1993), IASSNS-HEP-93/37.
\bibitem{DNII}M. Drees and M.M. Nojiri, MAD-PH-768 (1993).
\bibitem{Nojiri}M.M. Nojiri, private communication.
\bibitem{msw} L. Wolfenstein, Phys. Rev. D {\bf17} (1978) 2369; D {\bf20}
(1979) 2634; S. P. Mikheyev and A. Yu. Smirnov, Yad. Fiz. {\bf42} (1985) 1441;
Nuo. Cim. {\bf9C} (1986) 17.
\bibitem{erfm}J. Ellis, R. Flores, and S. Masood, Phys. Lett. B {\bf294} (1992)
229.
\bibitem{ssaw} T. Yanagida, Prog. Theo. Phys. {\bf B135} 66, 1978;
M. Gell-Mann, P. Ramond, and R. Slansky in {\em Supergravity},
ed. P. van Nieuwenhuizen and D. Freedman (North-Holland, Amsterdam 1979)
p. 315.
\bibitem{eln} J. Ellis, J.L. Lopez, and  D.V. Nanopoulos,
Phys. Lett. B {\bf 292} (1992) 189; J. Ellis, J.L. Lopez,
D.V. Nanopoulos and K. Olive, Phys. Lett. B {\bf 308} (1993) 70.

\end{thebibliography}
\end{document}